\documentclass[a4paper, twoside, 12pt]{article}

\usepackage[noblocks]{authblk} %% Affiliations with superscripts (1,2,3)
\usepackage[numbers]{natbib}
\usepackage{graphicx}
\usepackage{fancyhdr}
\usepackage[colorlinks=true, linkcolor=MidnightBlue, urlcolor=MidnightBlue, citecolor=PineGreen]{hyperref}
\usepackage[tableposition=top, justification=justified, textfont=it]{caption}
\usepackage{titlesec}
\usepackage{multicol}
\usepackage{afterpage}
\usepackage{lipsum}
\usepackage[dvipsnames, x11names]{xcolor}
\usepackage[inner=4cm, outer=4cm, top=2.8cm, bottom=2.8cm, marginparsep=0cm, marginparwidth=0cm, includehead, includefoot]{geometry}
\usepackage{setspace}
\usepackage{array}
\usepackage{bm}
\usepackage{tikz}
\usepackage{float}
\usepackage[framemethod]{mdframed}
\usepackage{amsmath}
\usepackage{amssymb}

\titleformat{\section}{\Large \bfseries \centering \scshape}{\thesection.}{0.3em}{}[{\titlerule[0.5pt]}]

\definecolor{shadecolor}{RGB}{230,230,230}
\newcommand{\mybox}[1]{\par\noindent\colorbox{shadecolor}
{\parbox{\dimexpr\textwidth-2\fboxsep\relax}{#1}}}

\titleformat{\subsection}{\large \bfseries \mybox}{\thesubsection}{1em}{}

\titleformat{\subsubsection}{\itshape}{\thesubsubsection.}{0.3em}{}

\renewenvironment{abstract}
{\vskip 2.5ex {\large\bf\noindent Abstract}\vspace{0.7ex} \\ %
  \bgroup\noindent\ignorespaces}%
{\par\egroup\vskip 2.5ex}

{\par\egroup\vskip 10ex}

\fancypagestyle{firstpage}{%
  
  \fancyhead[]{}
  \fancyfoot[C]{}
  \fancyfoot[L]{\footnotesize To appear in \\
  O. Colliot (Ed.), \textit{Machine Learning for Brain Disorders}, Springer\\
  }

}

\fancypagestyle{otherpages}{%
  
  \fancyhead[LE]{\thepage}
  \fancyhead[RE]{\runningauthor}
  \fancyhead[LO]{\runningheadtitle}
  \fancyhead[RO]{\thepage}
  \fancyfoot[L]{\footnotesize  \textit{Machine Learning for Brain Disorders}, Chapter~\chapternumber}
  \fancyfoot[C]{}
}

\makeatletter
\renewcommand{\maketitle}{\bgroup\setlength{\parindent}{0pt}

\begin{flushright}
  \color{MidnightBlue}
  \textbf{\LARGE Chapter~\chapternumber}
\end{flushright}

\vspace{0.3in}

\begin{flushleft}
    \setstretch{2.0} %% allows a nicer formatting of the title by adding more space between lines
    \textbf{\color{MidnightBlue}\huge\@title}
\end{flushleft}

\vspace{0.15in}

\begin{flushleft}
    \textbf{\bfseries \large\@author}
\end{flushleft}\egroup
}

\makeatother
\DeclareCaptionLabelFormat{labelformat}{\color{MidnightBlue}\bfseries #1 #2}
\renewcommand{\bibpreamble}{\scriptsize \begin{multicols}{2}}
\renewcommand{\bibpostamble}{\end{multicols}}

\mdfsetup{skipabove=\topskip, skipbelow=\topskip}

\newcounter{nicebox}
\newenvironment{nicebox}[1][]{%
    \refstepcounter{nicebox}%
    \ifstrempty{#1}%
    {\mdfsetup{%
        frametitle={%
            \tikz[baseline=(current bounding box.east),outer sep=0pt]
            \node[anchor=east,rectangle,fill=blue!20]
            {\strut Theorem~\thetheo};}}
    }%
    {\mdfsetup{%
        frametitle={%
            \tikz[baseline=(current bounding box.east),outer sep=0pt]
            \node[anchor=east,rectangle,fill=blue!20]
            {\strut Box~\thenicebox:~#1};}}%
    }%
    \mdfsetup{innertopmargin=10pt,linecolor=blue!20, linewidth=2pt,topline=true, frametitleaboveskip=\dimexpr-\ht\strutbox\relax,}
    \begin{mdframed}[]\relax%
    }{\end{mdframed}}

\newfloat{floatbox}{thp}{lop}
\floatname{floatbox}{Box}

\DeclareMathAlphabet{\mathsfit}{\encodingdefault}{\sfdefault}{m}{sl}
\SetMathAlphabet{\mathsfit}{bold}{\encodingdefault}{\sfdefault}{bx}{n}

\usepackage{graphicx}
\usepackage{verbatim}
\usepackage{caption}
\usepackage{subcaption}
\usepackage{doi}

\newcommand{\Sec}[1]{Section~\ref{#1}}

\newcommand{\DPM}{D$^{3}$PM}
\newcommand{\etal}{\textit{et al.}}

\begin{document}

%%%%%%%%%%%%%%%%%%%%%%%%%%%%%%%%%%%%%%%%%%%%%%%%%%%%%%%%%%%%%%%%%
%           I N F O R M A T I O N   T O   C H A N G E
%%%%%%%%%%%%%%%%%%%%%%%%%%%%%%%%%%%%%%%%%%%%%%%%%%%%%%%%%%%%%%%%%

% Author name displayed in the running head
\newcommand{\runningauthor}{Oxtoby}  %\textit{\etal.}

% Title displayed in the running head
\newcommand{\runningheadtitle}{Disease Progression Modelling}

% Chapter number
\newcommand{\chapternumber}{17}

% E-mail address of the corresponding author
\newcommand{\emailaddress}{n.oxtoby@ucl.ac.uk}

% Title of the chapter
\title{Data-Driven Disease Progression Modelling} 

% Authors' names and affiliation numbers
\author[*,1]{Neil P.~Oxtoby}

% Affiliations
\affil[1]{UCL Centre for Medical Image Computing, Department of Computer Science, University College London, London, UK}

%%%%%%%%%%%%%%%%%%%%%%%%%%%%%%%%%%%%%%%%%%%%%%%%%%%%%%%%%%%%%%%%%
\affil[*]{E-mail address: \href{mailto:\emailaddress}{\emailaddress} \href{https://disease-progression-modelling.github.io}{https://disease-progression-modelling.github.io}}

\maketitle

% Restore the geometry and change the page style for the other pages
\afterpage{\aftergroup\restoregeometry}
\pagestyle{otherpages}

% Abstract
\begin{abstract}
Intense debate in the Neurology community before 2010 culminated in hypothetical models of Alzheimer's disease progression: a pathophysiological cascade of biomarkers, each dynamic for only a segment of the full disease timeline. Inspired by this, data-driven disease progression modelling emerged from the computer science community with the aim to reconstruct neurodegenerative disease timelines using data from large cohorts of patients, healthy controls, and prodromal/at-risk individuals. This chapter describes selected highlights from the field, with a focus on utility for understanding and forecasting of disease progression.
\end{abstract}

% Keywords
%\begin{keywords}
%modelling, data, disease progression, keyword 4
%\end{keywords}

\section{Introduction}
\label{sec:intro}

Chronic progressive diseases are a major drain on social and economic resources. Many of these diseases have no treatments and no cure. In particular, age-related chronic diseases such as neurodegenerative diseases of the brain are a global healthcare pandemic-in-waiting as most of the world's population is living ever longer. A key example is Alzheimer's disease --- the leading cause of dementia --- but there are numerous other conditions that cause abnormal deterioration of brain tissue, leading to loss of cognitive performance, bodily function, independence, and ultimately death. Despite the increasing socioeconomic burden, neurodegenerative disease research has made impressive progress in the past decade, driven largely by the availability of large observational datasets and the computational analyses this enables.

Understanding neurodegenerative diseases is vital if they are to be managed, or even cured, but our understanding remains poor despite impressive progress in recent years. This poor understanding can be attributed to the many challenges of neurodegenerative diseases: no well-defined time axis due in part to heterogeneity in onset/speed/presentation, and censoring/attrition especially in later stages as patients deteriorate. These challenges, coupled with intense debate in the Neurology community (hypothetical models \cite{jack_hypothetical_2010,Aisen2010}) and increasing availability of data, piqued the interest of computational researchers aiming to provide quantitative answers to the mysteries of neurodegenerative diseases. This has ranged from vanilla off-the-shelf machine learning approaches through to more holistic statistical modelling approaches, the most advanced of which is data-driven disease progression modelling (\DPM).

\DPM{}s are defined by two key features: 1) they simultaneously reconstruct the disease timeline and estimate the quantitative disease signature/trajectory along this timeline; and 2) they are directly informed by observed data. \DPM{}s strike a balance between pure unsupervised learning, which requires truly big data, and traditional longitudinal modelling, which relies on a well-defined temporal axis --- neither of which are available in neurodegenerative diseases. For a review of the history and development of \DPM, see \cite{oxtoby_imaging_plus_x_2017}.

\begin{figure}[!htbp]
\centering
\includegraphics[width=\columnwidth,trim=90 30 0 0,clip]{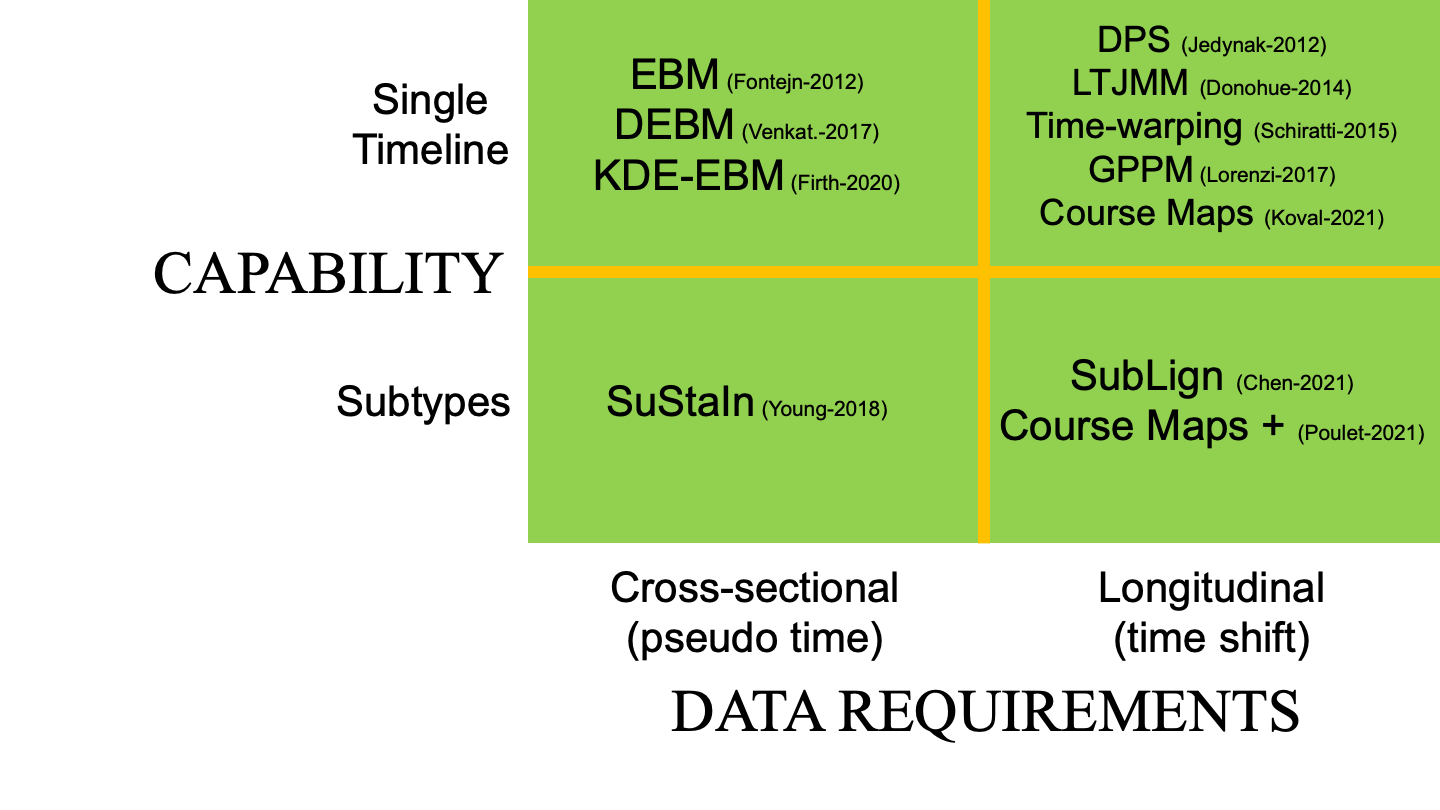}
\caption{\label{fig:capability}
Quadrant Matrix. \DPM{}s all estimate a disease timeline, with some capable of estimating multiple subtype timelines, using either cross-sectional data (pseudo-timeline) or longitudinal data (time-shift). \\Abbreviations: EBM -- Event Based Model; DEBM -- Discriminative EBM; KDE-EBM -- Kernel Density Estimation EBM; DPS -- Disease Progression Score; LTJMM -- Latent Time Joint Mixed Model; GPPM -- Gaussian Process Progression Model; SuStaIn -- Subtype and Stage Inference; SubLign -- Subtyping Alignment}
\end{figure}

The goal of this chapter is to highlight selected key \DPM{}s in a practical manner. The focus is on model capabilities and data requirements, aiming to inform the reader's \DPM{} analysis strategy based on the desired disease insight(s) and the data available. Figure \ref{fig:capability} places selected \DPM{}s on a Capability$\times$Data quadrant matrix: single timeline estimation vs subtyping, and cross-sectional vs longitudinal data availability. Table \ref{table:papers} lists more methodological papers relevant to \DPM, with model innovations grouped by the original paper for that method.

The chapter is organised as follows. It starts with a brief discussion of data preprocessing considerations in \Sec{sec:wrangling} --- an important step in medical data analysis. The treatment of \DPM{}s is separated into models for cross-sectional data (\Sec{sec:cross_sectional}) and models for longitudinal data (\Sec{sec:longitudinal}), each split into approaches that estimate a single timeline of disease progression and those capable of estimating multiple timelines within a dataset (subtyping). \Sec{sec:conclusion} concludes.

For a detailed timeline of \DPM{} development including taxonomy and pedigree of key models, see the Appendix.

\section{Data preprocessing}
\label{sec:wrangling}

This section briefly touches on two common preprocessing steps before fitting a \DPM{} to data from a progressive condition such as an irreversible chronic disease: controlling for confounding variables, and handling missing data. We refer to input features as biomarkers, and use ``covariate'' and ``confounder'' interchangeably. Missing data can refer to irregular/variable visits across individuals, or missing biomarker data due to one or more measurements not being performed for some reason. This section deals with the latter, since longitudinal models can typically handle irregular visits.

Controlling for confounding variables is an important element of any \DPM{} analysis. This helps to prevent the \DPM{} from learning non-disease-related patterns such as due to confounding covariates. Confounders can be included as covariates in certain models --- to account for that source of variation alongside other variables of interest. Another approach, often used for continuous-valued confounders, is to ``regress out'' this source of variation prior to fitting a model --- to remove non-disease-related signal in the data. This process involves training regression models on data from control participants (who are not expected to develop the disease being studied), then removing the relevant trends from all data. This method can also be applied to categorical risk factors (discrete variables). The canonical example of a potentially confounding variable in neurodegenerative diseases of the brain is age --- a key risk factor in many chronic diseases. Removing normal ageing signal is often phrased as ``adjusting for'' or ``controlling for'' age.

Handling missing data is an active area of research with a considerable body of literature. Broadly speaking there are two strategies. The easiest is to exclude participants having any missing biomarker (or covariate) data, but this can considerably reduce the sample size of data available for \DPM{} analysis. The second approach is to impute the missing data, e.g., using group mean values. Imputation can be explicit, or implicit. An example of implicit imputation is in Bayesian models that map data to probabilities, then deal with missing data probabilistically such as in the event-based model~\cite{young_media_2015} where $P(event | x)=0.5$ represents maximal uncertainty such as when a measurement $x$ is missing.

%Paragraphs:
%\begin{itemize}
% \item Scope: chronic progressive diseases. In brain: neurodegenerative diseases.
% \item The challenge of neurodegenerative diseases: no disease time axis, silent asymptomatic period, $\Rightarrow$ data scarcity, censoring, etc.
%\item Traditional ML cannot handle this because we don't have enough data.
%\item ``The journey to \DPM'' \cite{oxtoby_imaging_plus_x_2017}
%\end{itemize}

\section{Models for cross-sectional data}
\label{sec:cross_sectional}

\begin{nicebox}[Models for Cross-sectional data]
    \label{box:crosssectional}
    \begin{itemize}
        \item Pro: Data-economical. \\Require cross-sectional data only.
        \item Con: Limited forecasting utility. \\Forecasting requires augmentation with longitudinal data.
        \item Key application(s): assessing disease severity from a single visit, e.g., economical stratification for clinical research/trials.
    \end{itemize}
\end{nicebox}

\subsection{Single timeline estimation using cross-sectional data}
\label{sec:cross_sectional:single}

There is only one framework for estimating disease timelines from cross-sectional data: event-based modelling.

\subsubsection{Event Based Model}
\label{sec:ebm}

The event-based model (EBM) emerged in 2011 \cite{Fonteijn2011,Fonteijn2012}. The concept is simple: in a progressive disease, biomarker measurements only ever get worse, i.e., become increasingly and irreversibly abnormal. Thus, among a cohort of individuals at different stages of a single progressive disease, the cumulative sequence of biomarker abnormality events can be inferred from only a single visit per individual. This requires making a few assumptions: measurements from individuals are independent and represent samples from a single sequence of cumulative abnormality, i.e., a single timeline of disease progression. Such assumptions are commonplace in many statistical analyses of disease progression, and are reasonable approximations to make when analysing data from research studies that typically have strict inclusion and exclusion criteria to focus on a single condition of interest. Unsurprisingly, the event-based model has proven to be extremely powerful, producing insight into many neurodegenerative diseases: sporadic Alzheimer’s disease \cite{Young2014,oxtoby_frontiers_2017,kdeebm_firth_2020,Janelidze2021}, familial Alzheimer’s disease \cite{Fonteijn2012,Oxtoby2018}, Huntington’s disease \cite{Fonteijn2012,wijeratne_actn_2018}, Parkinson's disease \cite{oxtoby_brain_2021}, and others \cite{eshaghi_brain_2018,firth_ebm_downs_2018}.

%Mathematically, the event-based model for $N$ biomarker input features and $M$ individuals is expressed as the posterior probability of sequence $S$ (length $N$) of events $E$ given the data $X$ ($M$ rows, $N$ columns) via Bayes' theorem as $P(S|X) = P(S)P(X|S) / P(X)$ where the likelihood model is \cite{Fonteijn2012}
%\begin{equation}
%P(X|S) = \prod_{j=1}^J 
%\left\{
%\sum_{k=0}^N p(k) \left[ 
%\prod_{i=1}^k p(x_{i,j}|E)
%\prod_{i=k+1}^N p(x_{i,j}|\neg{E})
%\right]
%\right\}
%\label{eq:EBM}
%\end{equation}
%where $p(k)$ --- the prior probability of a measurement being at sequence position $k$ --- is used to integrate out the hidden/unknown variable $k$, and the pre-event/post-event likelihoods $p(x | E)$/$p(x | \neg{E})$ are determined by mixture modelling. For further details, see the EBM papers \cite{Fonteijn2012,Young2014,kdeebm_firth_2020}. In short, the marginal distribution $P(X)$ is analytically intractable so MCMC is used to sample from the posterior. For small $N\lesssim10$, it can be computationally feasible to perform an exhaustive search over all possible $N!$ sequences to find the maximum likelihood/\textit{a posteriori} solution.

\paragraph{EBM fitting}\paragraph{}

\begin{figure}[!htp]
\begin{center}
\includegraphics[width=0.8\columnwidth]{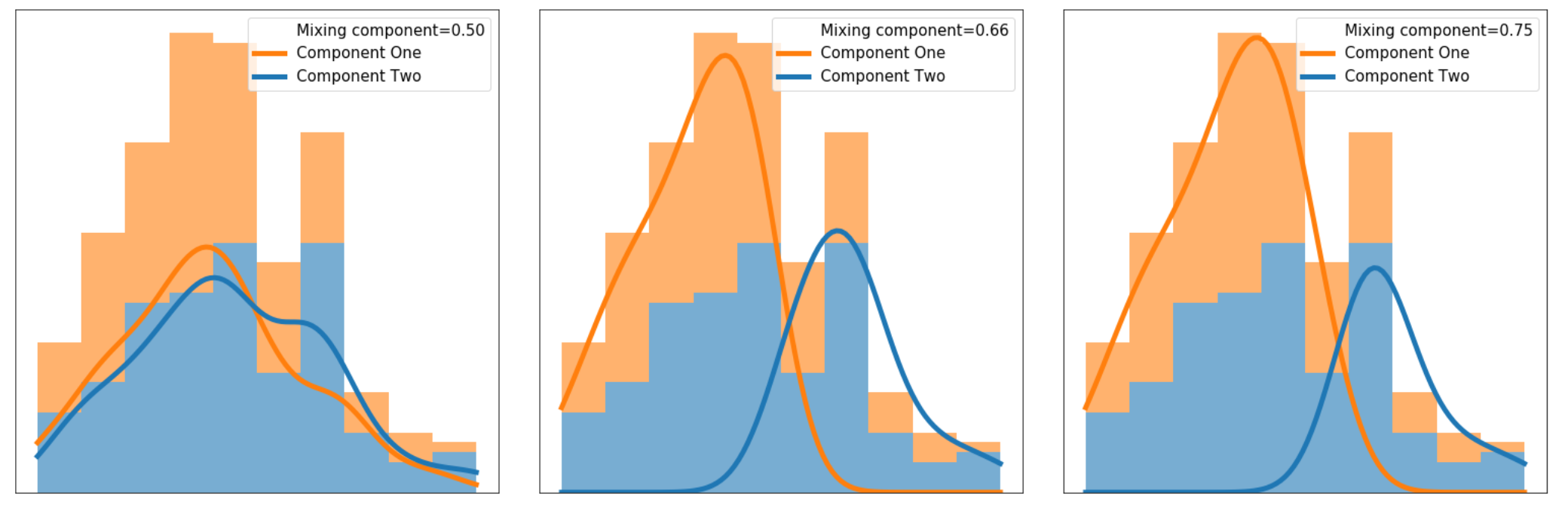}
\caption{\label{fig:mixturemodel}
Event-based models fit a mixture model to map biomarker values to abnormality probabilities. Left to right shows the convergence of a Kernel Density Estimate (KDE) mixture model. From Firth \etal., 2020 \cite{kdeebm_firth_2020} (CC BY 4.0).}
\end{center}
\end{figure}

The first step in fitting an event-based model maps biomarker values to abnormality values, similar to the hypothetical curves of biomarker abnormality proposed in 2010 \cite{jack_hypothetical_2010,Aisen2010}. The EBM does this probabilistically, using bivariate mixture modelling where individuals can be labelled either as pre-event/normal or post-event/abnormal to allow for (later) events that are yet to occur in patients, and similarly for the possibility of (earlier) events to have occurred in asymptomatic individuals. Various distributions have been proposed for this mixture modelling: combinations of uniform \cite{Fonteijn2011,Fonteijn2012}, Gaussian \cite{Fonteijn2011,Fonteijn2012,Young2014}, and kernel density estimate (KDE) distributions \cite{kdeebm_firth_2020}. This is visualised in Figure \ref{fig:mixturemodel}.

\begin{figure}[!htp]
\begin{center}
\includegraphics[width=0.8\columnwidth,trim=730 20 0 2,clip]{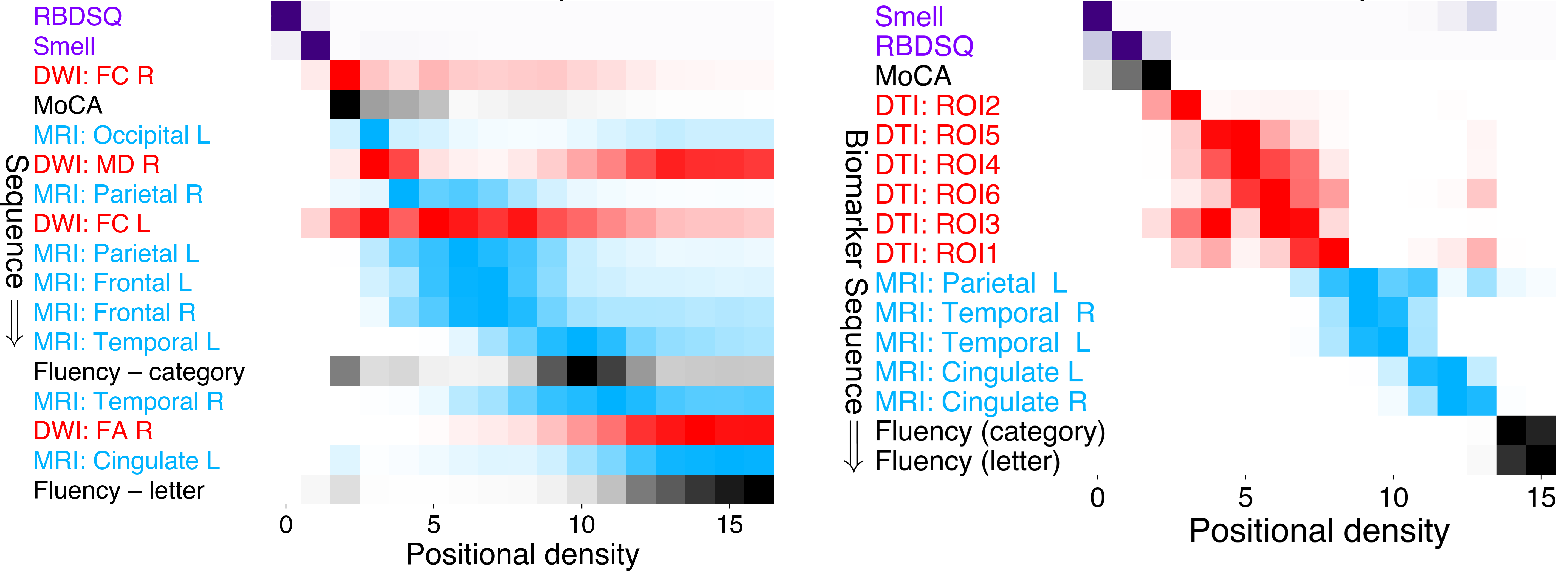}
\caption{\label{fig:pvd}
The event-based model posterior is a positional variance diagram showing uncertainty (left-to-right) in the maximum likelihood sequence (top-to-bottom). Parkinson's disease model from Oxtoby \etal., 2021 \cite{oxtoby_brain_2021} (CC BY 4.0).}
\end{center}
\end{figure}

The second step in fitting an EBM over $N$ events is to search the space of $N!$ possible sequences $S$ to reveal the most likely sequence (see \cite{Fonteijn2012,Young2014,kdeebm_firth_2020} for mathematical details). For small $N\lesssim10$, it can be computationally feasible to perform an exhaustive search over all possible $N!$ sequences to find the maximum likelihood/\textit{a posteriori} solution.
The EBM uses a combination of multiply-initialised gradient ascent, followed by MCMC sampling to estimate uncertainty in the sequence. This results in a model posterior that is a collection of samples from the posterior probability density for each biomarker as a function of sequence position. This is presented as a positional variance diagram~\cite{Fonteijn2012}, such as in Figure~\ref{fig:pvd}.

For further information and to try out EBM tutorials, the reader is directed to the open source \verb+kde_ebm+ package (\href{https://github.com/uckl-pond/kde_ebm}{github.com/ucl-pond/kde\_ebm}) and \href{https://disease-progression-modelling.github.io}{disease-progression-modelling.github.io}.

\subsubsection{Discriminative Event Based Model}
\label{sec:debm}

The discriminative event-based model (DEBM) was proposed in 2017 by Venkatraghavan \etal.~\cite{debm_ipmi_2017}. Whereas the EBM treats data from individuals as observations of a single group-level disease cascade (sequence), the DEBM estimates individual-level sequences and combines them into a group-level description of disease progression. This is done using a Mallow's Model, which is the ranking/sequencing equivalent of a univariate Gaussian distribution --- including estimation of a mean sequence and variance in this mean. Both EBM and DEBM estimate group-level biomarker abnormality using mixture modelling and both approaches directly estimate uncertainty in the sequence.

Additionally, Venkatraghavan \etal.~\cite{debm_ipmi_2017,Venkatraghavan_NIMG_2019} also introduced a pseudo-temporal ``disease time'' that converts the DEBM posterior into a continuous measure of disease severity.

\paragraph{DEBM fitting}\paragraph{}

As with the EBM, DEBM model fitting starts with mixture modelling (see \Sec{sec:ebm}). Next, a sequence is estimated for each individual by ranking the abnormality probabilities in descending order. A group-level mean sequence (with variance) is estimated by fitting the individual sequences to a Mallow's Model. For details, see \cite{debm_ipmi_2017,Venkatraghavan_NIMG_2019} and subsequent innovations to the DEBM. Notably, DEBM is often quicker to fit than EBM, which makes it appealing for high-dimensional extensions, e.g., aiming to estimate voxel-wise atrophy signatures from cross-sectional brain imaging data. % \cite{VikramsOtherDEBMPapers_CoInit_etc}.

For further information and to try it out, the reader is directed to the open source \verb+pyebm+ package (\url{https://github.com/88vikram/pyebm}).

\subsection{Subtyping using cross-sectional data}
\label{sec:cross_sectional:subtyping}

\begin{nicebox}[Subtyping models]
    \label{box:subtyping}
    \begin{itemize}
        \item Pro: Uncovering heterogeneity without conflating severity with subtype. \\Evidence suggests that disease subtypes exist.
        \item Con: Overly simplistic. \\Current models ignore comorbidity.
%        \item Con: requires large data. \\Cannot detect patterns that aren't sampled.
    \end{itemize}
\end{nicebox}

Augmenting the event-based model concept with unsupervised machine learning, Subtype and Stage Inference (SuStaIn) was introduced by Young \textit{\etal.}~\cite{young_sustain_2018-short}. This marriage of clustering to disease progression modelling has proven very powerful and popular, with high-impact results appearing in prominent journals for multiple brain diseases~\cite{vogel_natmed_2021,eshaghi_mssustain_2021,collij_amyloidsustain_2022}, chronic lung disease~\cite{young_copdsustain_2020}, and knee osteoarthritis~\cite{li_kneesustain_2021}. SuStaIn's popularity is perhaps unsurprising given that it was the first method capable of unravelling spatiotemporal heterogeneity (pathological severity across an organ) from phenotypic heterogeneity (disease subtypes) in progressive conditions using only cross-sectional data.

\begin{figure}[!htp]
\begin{center}
\includegraphics[width=0.9\columnwidth]{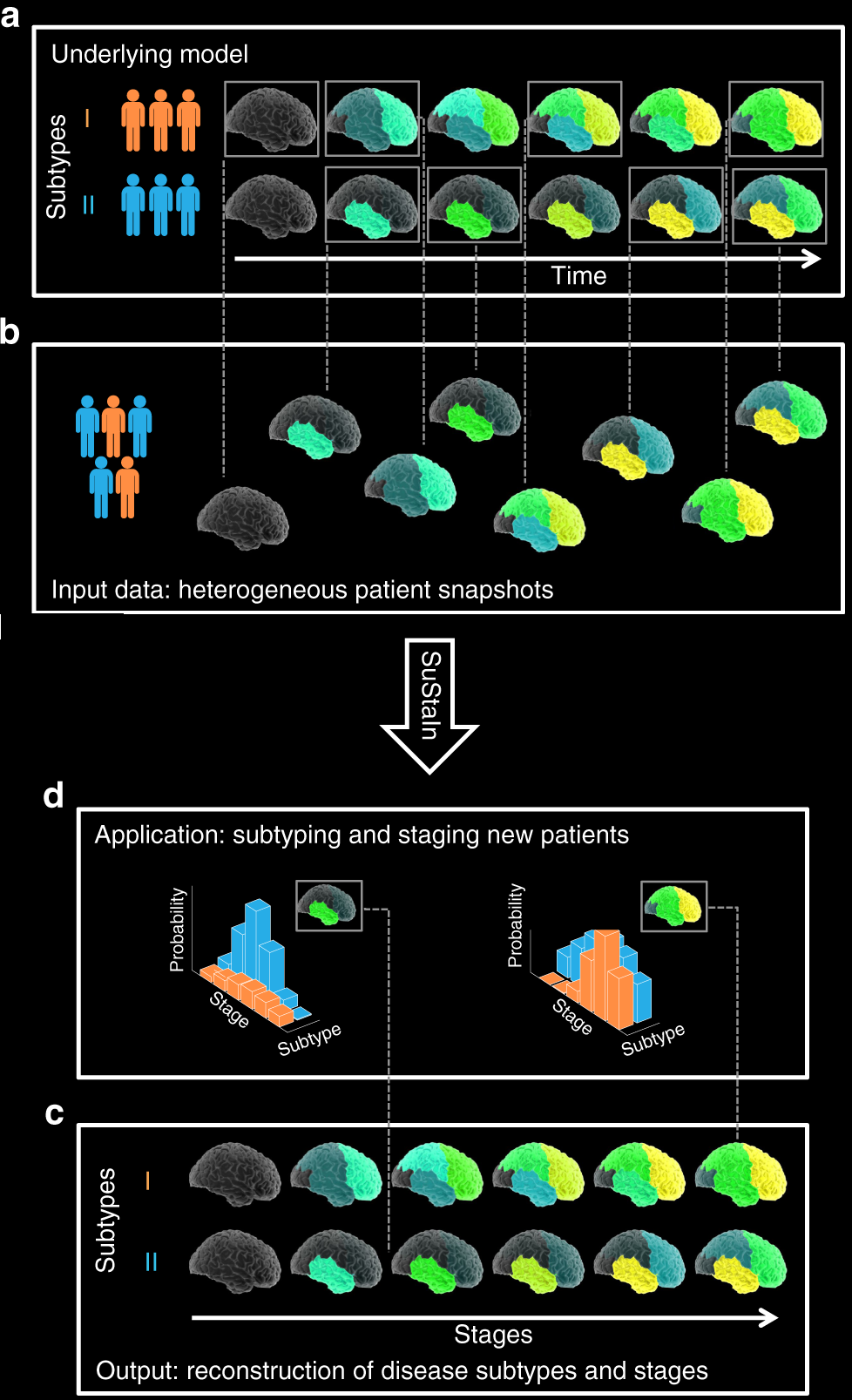}
\caption{\label{fig:sustain}
The concept of Subtype and Stage Inference (SuStaIn). Reproduced from Young \textit{\etal.}, 2018 \cite{young_sustain_2018-short} (CC BY 4.0).}
\end{center}
\end{figure}

Figure \ref{fig:sustain} (adapted from~\cite{young_sustain_2018-short}) shows the concept behind SuStaIn. SuStaIn iteratively solves the clustering problem from 1 to $N_\mathrm{S}^\mathrm{max}$ subtypes. The $N_\mathrm{S}$ model is fitted by splitting each of the $N_\mathrm{S}-1$ subtypes into two clusters then solving the $N_\mathrm{S}$-cluster problem, which produces $N_\mathrm{S}-1$ candidate $N_\mathrm{S}$-cluster models, from which the maximum likelihood model is chosen, then the algorithm continues to $N_\mathrm{S}+1$ and so on.

Young \textit{\etal.}~\cite{young_sustain_2018-short} also introduced the z-score event progression model that breaks down individual biomarker events into piecewise linear transitions between z-scores of interest. This removes the need for mixture modelling (such as in event-based modelling) and enables inference to be performed at sub-threshold biomarker values.

\paragraph{SuStaIn fitting}\paragraph{}

For the user, a SuStaIn analysis is very similar to an event-based model analysis. For further information, the reader is directed to the open source \verb+pySuStaIn+ package~\cite{pySuStaIn} (\url{https://github.com/ucl-pond/pySuStaIn}), which includes tutorials. As well as the z-score progression model, \verb+pySuStaIn+ includes the various event-based models (see \Sec{sec:cross_sectional:single}), and the more recent scored-events model for ordinal data~\cite{young_ordinal_sustain_2021} such as visual ratings of medical images.

\section{Models for longitudinal data}
\label{sec:longitudinal}

\begin{nicebox}[Models for Longitudinal data]
    \label{box:longitudinal}
    \begin{itemize}
        \item Pro: Good forecasting utility. \\High temporal precision allows individualised forecasting.
        \item Con: Data-heavy. \\Require longitudinal data (multiple visits, years). Can be slow to fit.
        \item Key application(s): assessing speed of disease progression; assessing individual variability.
    \end{itemize}
\end{nicebox}

The availability of longitudinal data has fuelled development of more sophisticated \DPM{}s, inspired by mixed models. Mixed (effect) modelling is the workhorse of longitudinal statistical analysis against a known timeline, e.g., age. Mixed models provide a hierarchical description of individual-level variation (random effects) about group-level trends (fixed effects), hence the common parlance "mixed effects" models. Many of the \DPM{}s for longitudinal data discussed below are in fact mixed models with an additional latent-time parameter that characterises the disease timeline. Similar approaches in various fields are known as ``self-modelling regression'' or ``pseudo-time'' models. We focus on parametric models, but also mention nonparametric models, and an emerging hybrid discrete-continuous model.

\subsection{Single timeline estimation using longitudinal data}
\label{sec:longitudinal:single}

There are both parametric and non-parametric approaches to estimating disease timelines from longitudinal data. The common goal is to ``stitch together'' a full disease timeline (decades long) out of relatively short samples from individuals (a few years each) covering a range of severity in symptoms and biomarker abnormality. Some of the earliest work emerged from the medical image registration community, where ``warping'' images to a common template is one of the first steps in group analyses~\cite{durrleman_miccai_2009}.

Broadly speaking, there are two categories of \DPM{}s for longitudinal data: time-shifting models and differential equation models. Time-shifting models translate/deform the individual data, metaphorically stitching them together into a quantitative template of disease progression. Differential equation models estimate a statistical model of biomarker dynamics in phase-plane space (position vs velocity), which is subsequently inverted to produce biomarker trajectories.

\subsubsection{Explicit models for longitudinal data: latent-time models}
\label{sec:long_timeshift}

Jedynak, \etal\ \cite{jedynak_computational_2012} introduced the Disease Progression Score (DPS) model in 2012, which aligns biomarker data from individuals to a group template model using a linear transformation of age into a disease progression score $s_i=\alpha_i \mathrm{age} + \beta_i$. Individuals have their own rate of progression $\alpha_i$ (constant over the short observation time) and disease onset $\beta_i$. Group level biomarker dynamics are modelled as sigmoid (``S'') curves. A Bayesian extension of the DPS approach (BPS) appeared in 2019 \cite{BilgelDADM2019}. Code for both the DPS and BPS was released publicly: \url{https://www.nitrc.org/projects/progscore}; \url{https://hub.docker.com/r/bilgelm/bayesian-ps-adni/}.

Donohue, \etal\ \cite{Donohue2014} introduced a self-modelling regression approach similar to the DPS model in 2014. It was later generalised into the more flexible Latent Time Joint Mixed (effects) Model (LTJMM) \cite{li_ltjmm_2019}, which can include covariates as fixed effects and is a flexible Bayesian framework for inference. The LTJMM software was released publicly to \url{https://bitbucket.org/mdonohue/ltjmm}.

A non-parametric latent-time mixed model appeared in 2017: the Gaussian process progression model (GPPM) of Lorenzi, \etal\ \cite{lorenzi_gppm_2019}. This is a flexible Bayesian approach akin to (parametric) self-modelling regression that doesn't impose a parametric form for biomarker trajectories. More recent work supplemented the GPPM with a dynamical systems model of molecular pathology spread through the brain \cite{garbarino_GPPM-DS_2021} that can regularise the GPPM fit to produce a more accurate disease timeline reconstruction that also provides insight into neurodegenerative disease mechanisms (which could be a standalone chapter of this book). The GPPM and GPPM-DS model source code was released publicly via \href{https://gitlab.inria.fr/epione/GP_progression_model_V2}{gitlab.inria.fr/epione} and tutorials are available at \href{https://disease-progression-model.github.io}{disease-progression-model.github.io}.

In 2015, Schiratti \etal\ \cite{schirattiNIPS2015,schiratti_ipmi_2015,SchirattiJMLR2017} introduced a general framework for estimating spatiotemporal trajectories for any type of manifold-valued data. The framework is based on Riemannian geometry and a mixed-effects model with time reparametrisation. It was subsequently extended by Koval et al \cite{Koval2021} to form the Disease Course Mapping approach (available in the \href{https://gitlab.com/icm-institute/aramislab/leaspy}{leaspy} software package). Disease Course Mapping combines time warping (of age) and inter-biomarker spacing translation. Time warping changes disease progression dynamics --- time shift/onset, and acceleration/progression speed --- but not the trajectory. Inter-biomarker spacings shift an individual's trajectory to account for individual differences in the timing and ordering of biomarker trajectories.

\begin{figure}[htbp]
\begin{subfigure}[t]{\textwidth}
\centering\includegraphics[width=\columnwidth,trim=0 0 0 0,clip]{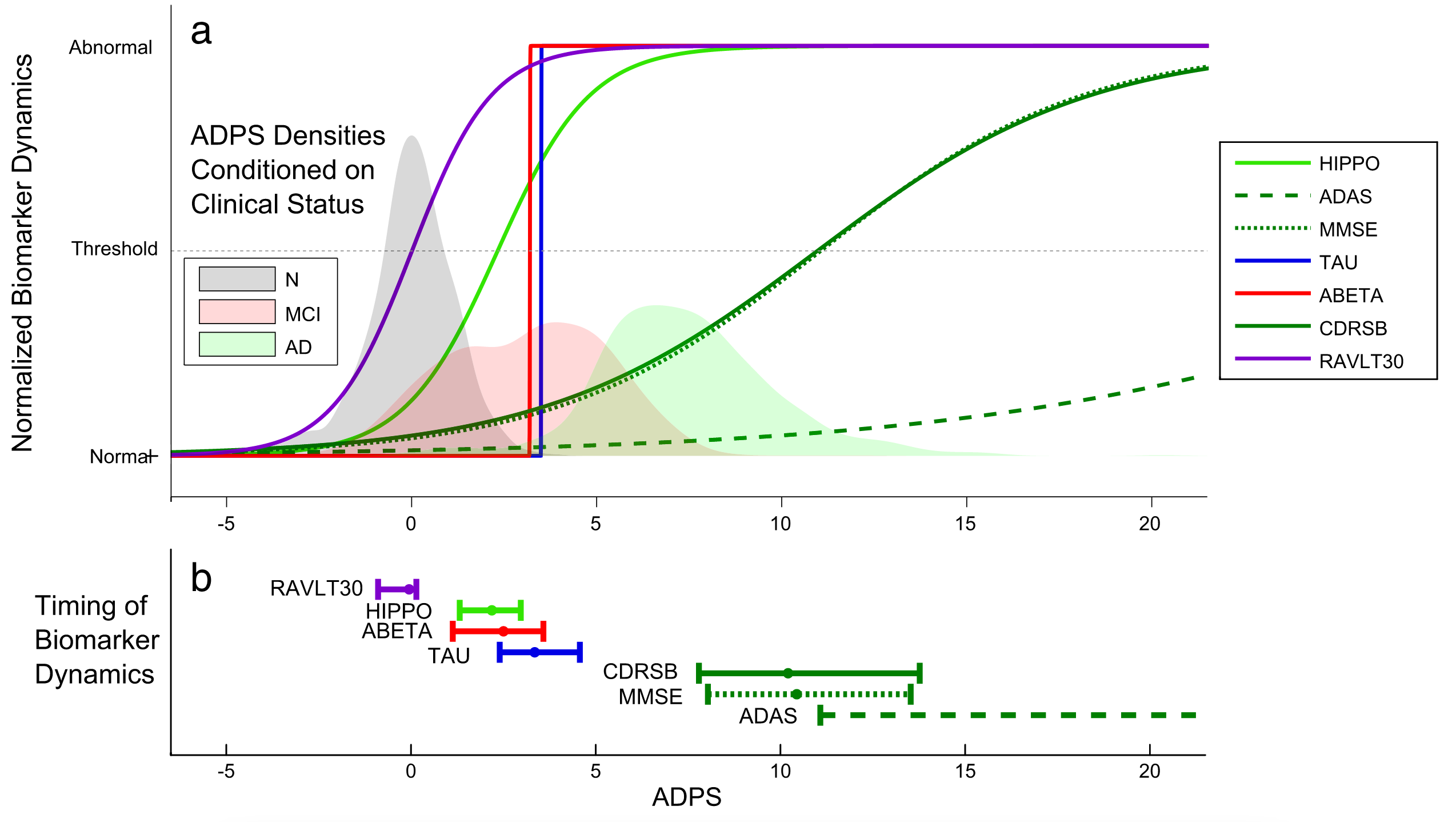}
\caption{Alzheimer's Disease Progression Score (2012) \cite{jedynak_computational_2012}. Reprinted from NeuroImage, Vol 63, Jedynak \etal, A computational neurodegenerative disease progression score: Method and results with the Alzheimer's Disease Neuroimaging Initiative cohort, 1478–1486, \textcopyright (2012), with permission from Elsevier.}
\label{fig:dps}
\end{subfigure}
\begin{subfigure}[t]{\textwidth}
\centering\includegraphics[width=\columnwidth,trim=0 0 0 0,clip]{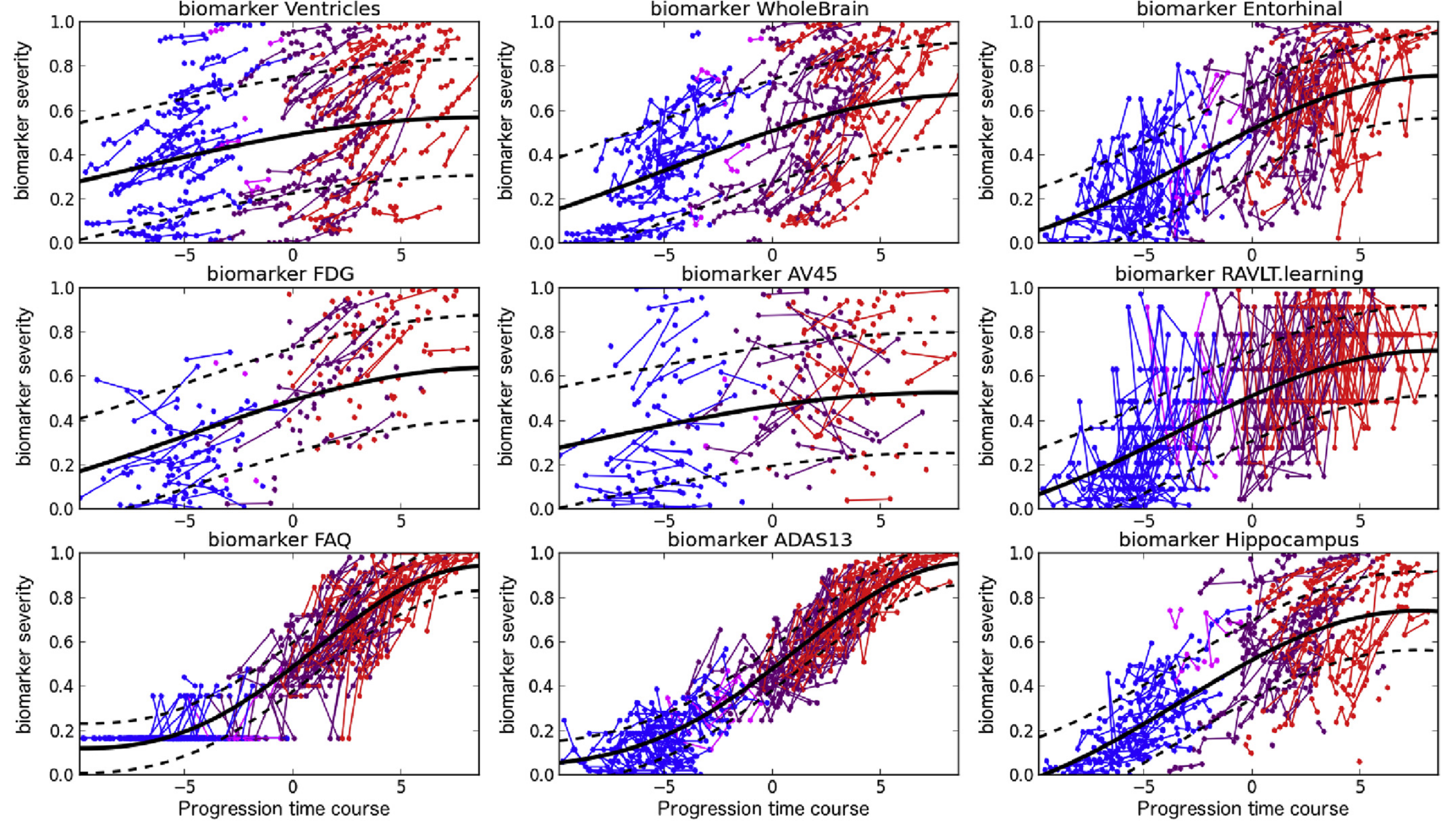}
\caption{Gaussian Process Progression Model (2017) \cite{lorenzi_gppm_2019}. Reprinted from NeuroImage, Vol 190, Lorenzi \etal, Probabilistic disease progression modeling to characterize diagnostic uncertainty: Application to staging and prediction in Alzheimer's disease, 56–68, \textcopyright (2019), with permission from Elsevier.}
\label{fig:gppm}
\end{subfigure}
\caption{Two examples of \DPM{}s fit to longitudinal data: Disease Progression Score \cite{jedynak_computational_2012} and Gaussian Process Progression Model \cite{lorenzi_gppm_2019}.\label{longitudinal1}}
\end{figure}

\begin{figure}[htbp]
\begin{subfigure}[t]{\textwidth}
\centering\includegraphics[width=\columnwidth,trim=0 0 0 0,clip]{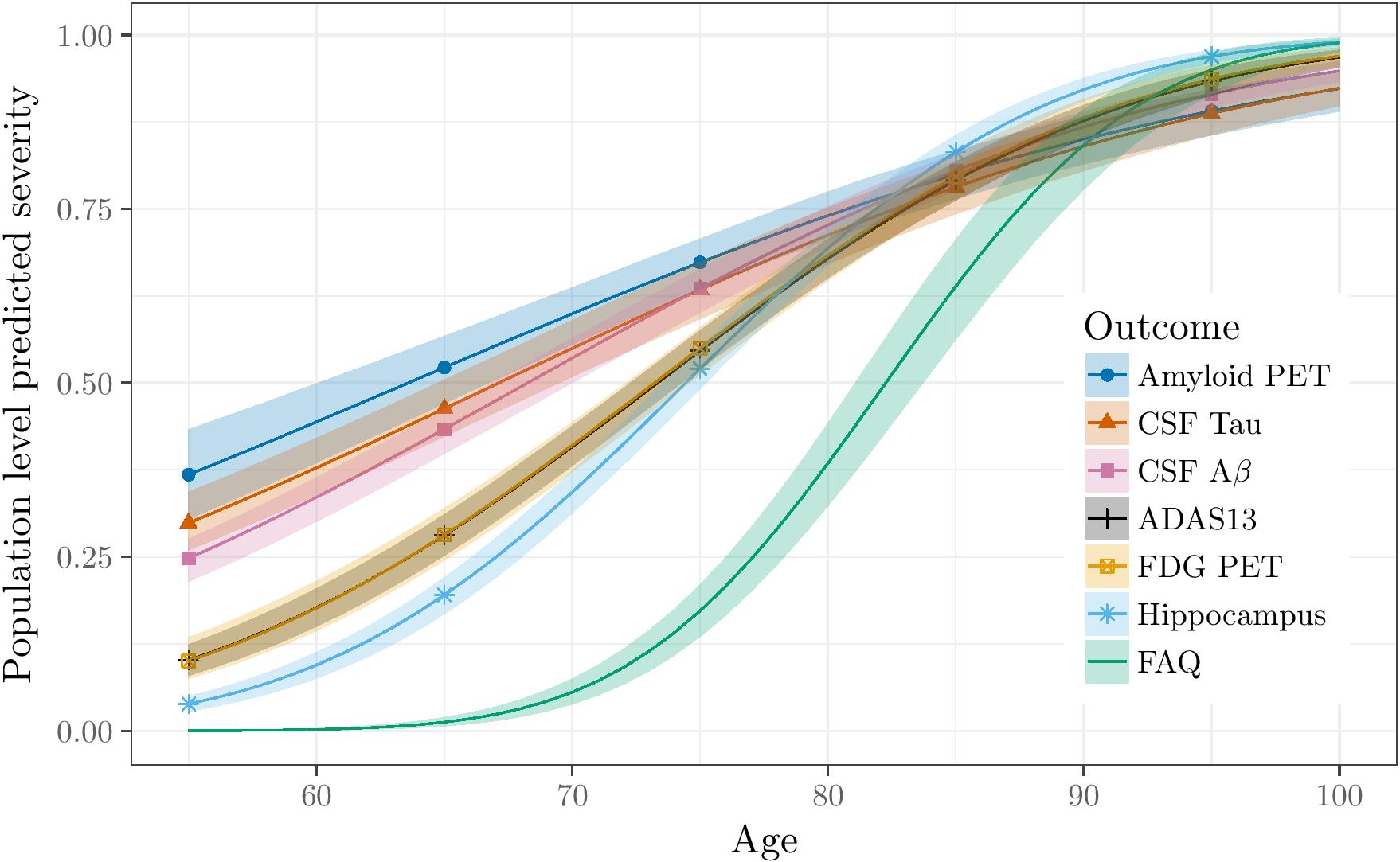}
\caption{Latent Time Joint Mixed Model (2017) \cite{li_ltjmm_2019}. From \cite{li_arxiv_2017} (CC BY 4.0).}
\label{fig:ltjmm}
\end{subfigure}
\begin{subfigure}[t]{\textwidth}
\centering\includegraphics[width=\columnwidth,trim=0 0 0 0,clip]{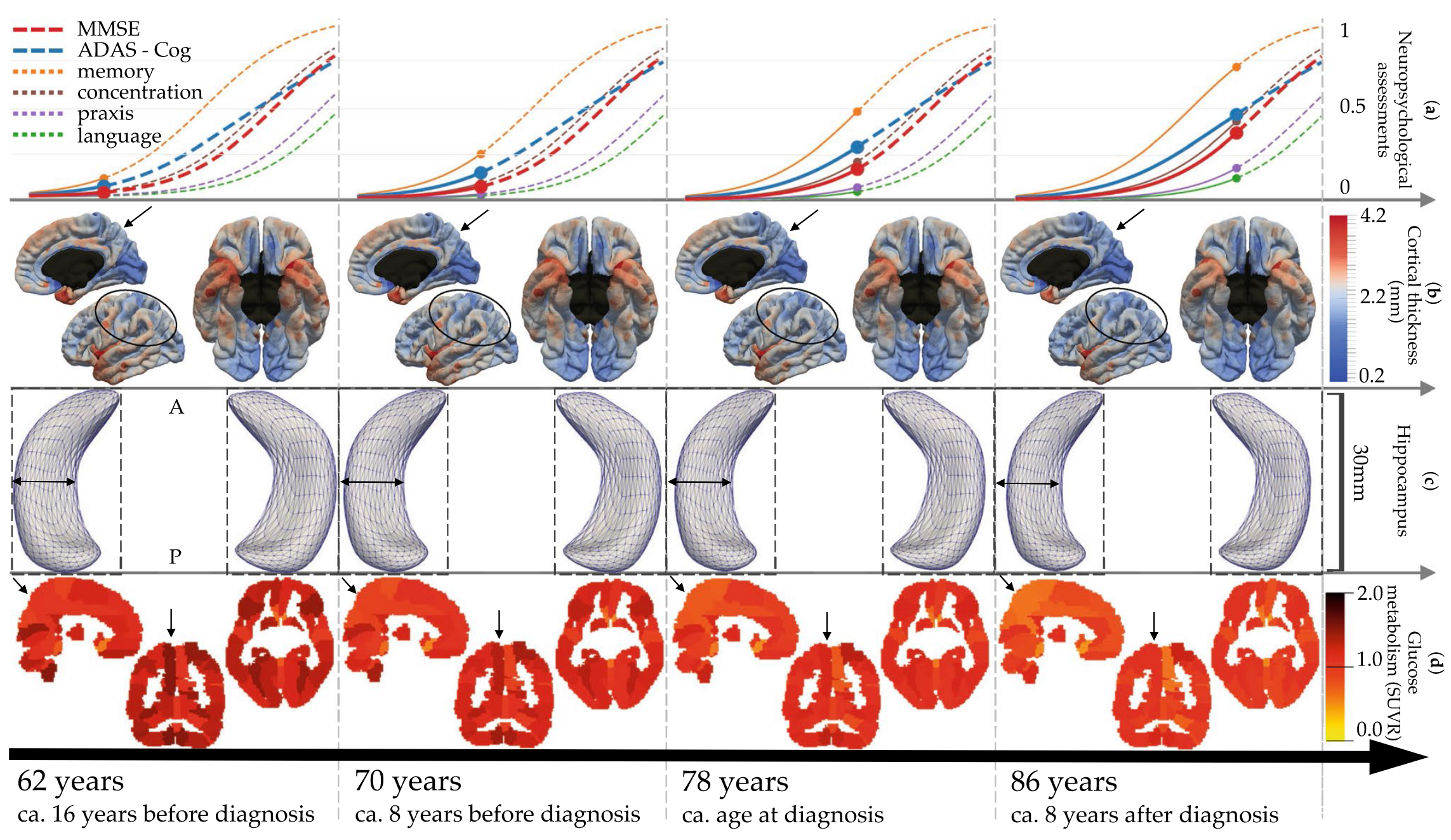}
\caption{Alzheimer's Disease Course Map (2021) \cite{Koval2021} (CC BY 4.0).}
\label{fig:dcm}
\end{subfigure}
\caption{Two additional examples of \DPM{}s fit to longitudinal data: Latent Time Joint Mixed Model \cite{li_ltjmm_2019} and Disease Course Mapping \cite{Koval2021}.\label{longitudinal2}}
\end{figure}

Figures \ref{longitudinal1} and \ref{longitudinal2} shows example outputs of these models when trained on data from older people at risk of Alzheimer's disease, including those with diagnosed mild cognitive impairment and dementia due to probable Alzheimer's disease.

\paragraph{Fitting longitudinal latent-time models}\paragraph{}
% Longitudinal biomarker data is commonly structured in long format where each row corresponds to a single time-point from a single individual.
Fitting \DPM{}s for longitudinal data is more complex than for cross-sectional data, and the software packages discussed above each expect the data in slightly different formats. One thing they have in common is that renormalisation (e.g., min-max or z-score) and reorientation (e.g., to be increasing) is required to put biomarkers on a common scale and direction. In some cases such preprocessing is necessary to ensure/accelerate model convergence. For example, the LTJMM used a quantile transformation followed by inverse Gaussian quantile function to put all biomarkers on a Gaussian scale. For further detailed discussion, including model identifiability, we refer the reader to the original publications cited above and the didactic resources at \href{https://disease-progression-modelling.github.io}{disease-progression-modelling.github.io}.

\subsubsection{Implicit models for longitudinal data: differential equation models}
\label{sec:long_differential}

Parametric differential equation \DPM{}s emerged between 2011--2014 \cite{Sabuncu2011,Samtani2012,Villemagne2013,Oxtoby2014}, receiving a more formal treatment in 2017 \cite{Budgeon2017}. In a hat-tip to Physics, these have also been dubbed ``phase-plane'' models, which aids in their understanding as a model of velocity (biomarker progression rate) as a function of position (biomarker value). Model fitting is a two-step process whereby the long-time biomarker trajectory is estimated by integrating the phase-plane model estimated on observed data.

A nonparametric differential equation \DPM{} using Gaussian Processes (GP-DEM) was introduced in 2018 \cite{Oxtoby2018}. This added flexibility to the preceding parametric approaches and produced state of the art results in predicting symptom onset in familial Alzheimer's disease.

\begin{figure}[htbp]
\centering
\includegraphics[width=0.98\columnwidth]{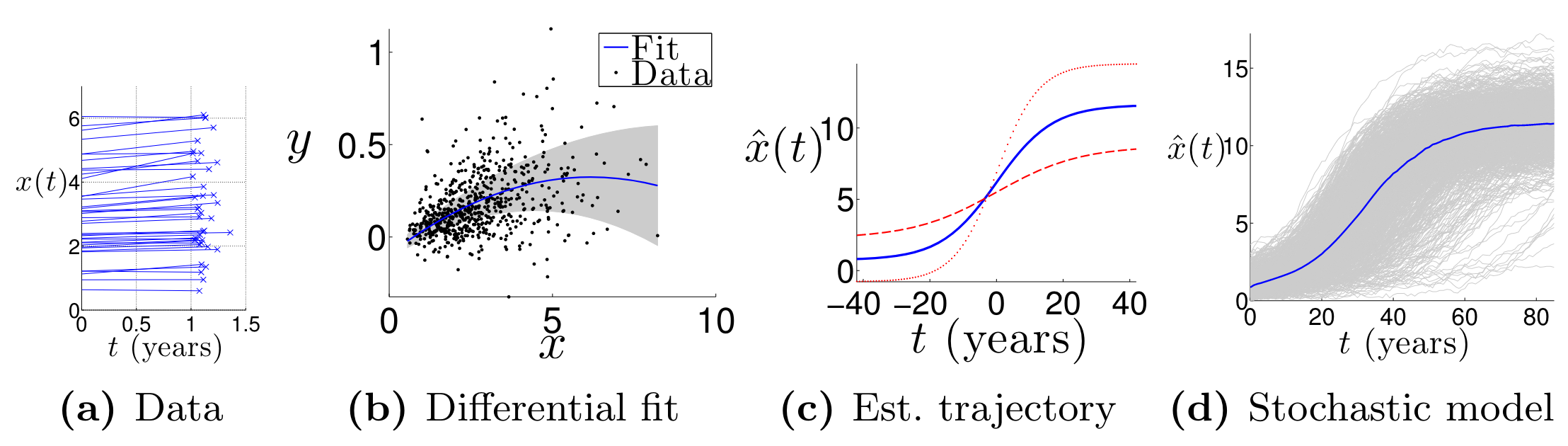}
\caption{Differential equation models, or phase-plane models, for biomarker dynamics involve a three step process: estimate individual-level position and velocity; fit a group-level model of velocity $y$ vs position $x$; integrate to produce a trajectory $x(t)$. \\Reprinted by permission from Springer Nature: Oxtoby, N.P. \etal, Learning Imaging Biomarker Trajectories from Noisy Alzheimer's Disease Data Using a Bayesian Multilevel Model. In: Cardoso, M.J., Simpson, I., Arbel, T., Precup, D., Ribbens, A. (eds) Bayesian and grAphical Models for Biomedical Imaging. Lecture Notes in Computer Science, vol 8677, pp. 85–94 \textcopyright (2014) \cite{Oxtoby2014}.\label{dem}}
\end{figure}

\paragraph{Fitting differential equation models}\paragraph{}
The concept is shown in Figure \ref{dem}: differential equation model fitting is a three-step process. First estimate a single value per individual of biomarker ``velocity'' and ``position'', then estimate a group-level differential equation model of velocity $y$ as a function of position $x$, which is integrated/inverted to produce a biomarker trajectory $x(t)$. For example, linear regression can produce estimates of position (e.g., intercept) and velocity (e.g. gradient). Differential equation models can be univariate or multivariate and can include covariates explicitly.

\subsubsection{Hybrid discrete-continuous models}
\label{sec:TEBM}

\begin{figure}[htbp]
\centering
\includegraphics[width=0.98\columnwidth]{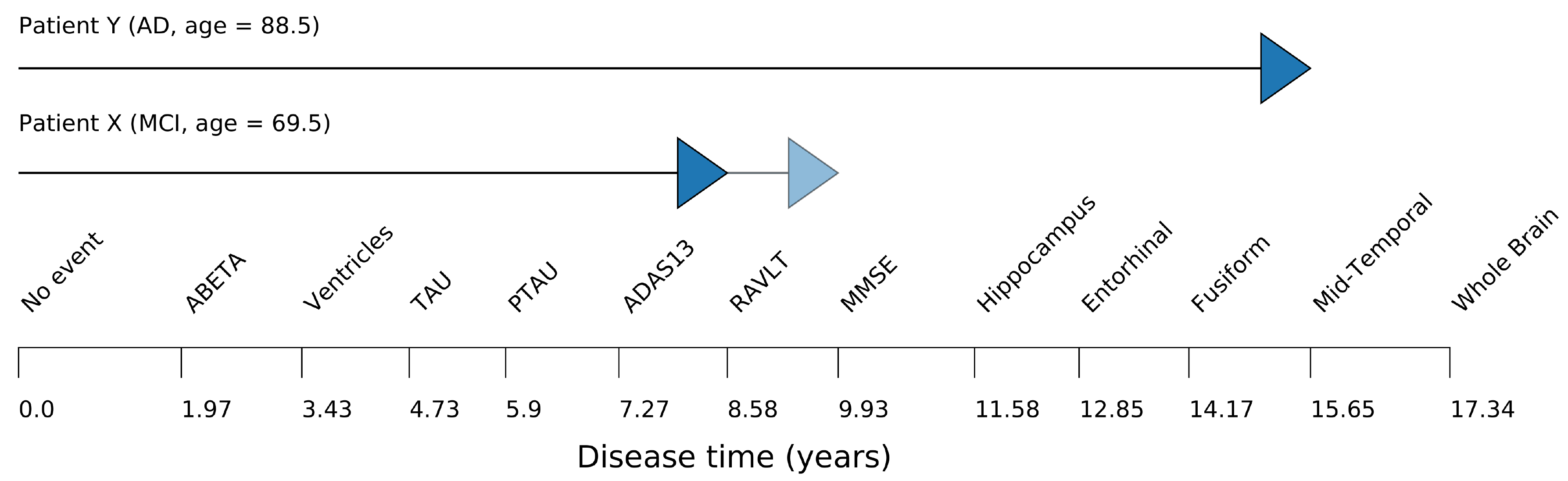}
\caption{Alzheimer's disease sequence and timeline estimated by a hybrid discrete-continuous \DPM: the Temporal Event-Based Model  \cite{wijeratne_learning_2020,wijeratne_TEBM_2021}. Permission to reuse was kindly granted by the authors of \cite{wijeratne_learning_2020}. \label{tebm}}
\end{figure}

Recent work introduced the Temporal EBM (TEBM) \cite{wijeratne_learning_2020,wijeratne_TEBM_2021}, which augments event-based modelling with hidden Markov modelling to produce a hybrid discrete-continuous \DPM. This is a halfway house between discrete models (great for medical decision making) and continuous models (great for detailed understanding of disease progression). Trained on data from ADNI, the TEBM revealed the full timeline of the pathophysiological cascade of Alzheimer's disease, as shown in Figure \ref{tebm}.

\subsection{Subtyping using longitudinal data}
\label{sec:longitudinal:subtyping}

Clustering longitudinal data without a well-defined time axis can be extremely difficult. Jointly estimating latent time for multiple trajectories is an identifiability challenge, i.e., multiple parameter combinations can explain the same data. This is particularly challenging when observations span a relatively small fraction of the full disease timeline, as in age-related neurodegenerative diseases.

Chen \etal{} \cite{sublign} introduced SubLign for subtyping and aligning longitudinal disease data. The authors frame the challenge eloquently as having misaligned, interval-censored data: left-censoring from patients being observed only after disease onset; right censoring from patient dropout in more severe disease. SubLign combines a deep generative model (based on a recurrent neural network \cite{Rumelhart1986}) for learning individual latent time-shifts and parametric biomarker trajectories using a variational approach, followed by k-means clustering. It was applied to data from a Parkinson's disease cohort to recover some known clinical phenotypes in new detail.

Poulet and Durrleman \cite{poulet_mixture_2021} recently added mixture-model clustering to the non-linear mixed model approach of Disease Course Mapping \cite{Koval2021}. The framework jointly estimates model parameters and subtypes using a modification of the Expectation-Maximisation algorithm. In simulated data experiments their approach outperforms a naive baseline. Experiments on real data in Alzheimer's disease distinguished rapid from slow clinical progression, with minimal differences in biomarker trajectories.

\section{Conclusion}
\label{sec:conclusion}

21st century Medicine faces many challenges due to ageing populations worldwide, including increasing socioeconomic burden from age-related brain disorders like Alzheimer's disease. Many failed clinical trials fuelled intense debate in Neurology in the first decade of this century, culminating in the prominent hypothesis of Alzheimer's disease progression as a pathophysiological cascade of dynamic biomarker events. This inspired the emergence of Data-Driven Disease Progression Modelling (\DPM) from the Computer Science community during the second decade of the 21st century --- an explosion of quantitative models for neurodegenerative disease progression enabling numerous high-impact insights across multiple brain disorders. The community continues to build and share open source code (see Box \ref{box:code}) and run machine learning challenges \cite{marinescu_TADPOLE_2018,marinescu_TADPOLE_2021-short,bron_ten_2022}. What will the third decade of the 21st century bring for this exciting subset of machine learning for brain disorders?

% Acknowledgments section
\section*{Acknowledgments}
The author is a UKRI Future Leaders Fellow (MR/S03546X/1) and acknowledges conversations and collaborations with colleagues from the UCL POND group (\url{http://pond.cs.ucl.ac.uk}), the EuroPOND consortium (\url{http://europond.eu}), the E-DADS consortium (\url{https://e-dads.github.io}; MR/T046422/1; EU JPND), and the open Disease Progression Modelling Initiative (\url{https://disease-progression-modelling.github.io}) %(\href{https://disease-progression-modelling.github.io}{disease-progression-modelling.github.io}) 
--- This project has received funding from the European Union’s Horizon 2020 research and innovation programme under grant agreement No.~666992.

\newpage
\section*{Appendix}

A taxonomy and pedigree of key \DPM{} papers is given in Table~\ref{table:papers}. Box \ref{box:code} contains links to open source code for \DPM{}s.

\begin{table}[htp]
\small\caption{A taxonomy and pedigree of \DPM{} papers\label{table:papers}. *Asterisks denote models for cross-sectional data.}%
\begin{center}
\begin{tabular}{| p{0.62\linewidth} | p{0.38\linewidth} |}
\hline
Reference (first author only) & Description \\
\hline \hline
Ashford, Curren. Psych. Rep. (2001) \cite{Ashford2001} & Differential equation \\ 
\hline
% Yang, \etal., JAD (2011) & \\ PDF seems corrupted!
Gomeni, Alz Dem (2011) \cite{gomeni_2011} & Differential equation \\ 
\hline
Sabuncu, Arch. Neurol. (2011) \cite{Sabuncu2011} & Differential equation \\ 
\hline
Samtani, J. Clin. Pharmacol. (2012) \cite{Samtani2012} & Differential equation \\ 
\hline
Jedynak, NIMG (2012) \cite{jedynak_computational_2012} & Progression score (linear) \\
$\Rightarrow$ Bilgel, IPMI (2015) \cite{bilgel_temporal_2015} &  \\ 
$\Rightarrow$ Bilgel, NIMG (2016) \cite{bilgel_multivariate_2016} & Latent time mixed effects \\
$\Rightarrow$ Bilgel, Alz Dem DADM (2019) \cite{BilgelDADM2019} & Bayesian \\ 
\hline
*Fonteijn, IPMI (2011) \cite{Fonteijn2011}; NIMG (2012) \cite{Fonteijn2012} & Cumulative events \\
$\Rightarrow$ *Young, Brain (2014) \cite{Young2014} & Robust for sporadic disease \\ 
$\Rightarrow$ *Venkatraghavan, IPMI (2017) \cite{debm_ipmi_2017}; NIMG (2019) \cite{Venkatraghavan_NIMG_2019} & Individual-level \\ 
$\Rightarrow$ *Young, Nat Commun (2018) \cite{young_sustain_2018-short} & +Subtyping, +Z-score Model \\
$\Rightarrow$ *Young, Frontiers (2021) \cite{young_ordinal_2021} & +Scored Events Model \\
$\Rightarrow$ *Firth, Alz Dem (2020) \cite{kdeebm_firth_2020} & +Nonparametric events \\ 
$\Rightarrow$ Wijeratne, ML4H2020, IPMI (2021) \cite{wijeratne_learning_2020,wijeratne_TEBM_2021} & +Transition times \\ 
\hline
Villemagne, Lancet Neurol (2013) \cite{Villemagne2013} & Differential equation \\
$\Rightarrow$ Budgeon, Stat. in Med. (2017) \cite{Budgeon2017} & Formalism \\ 
\hline
Durrleman, Int. J. Comput. Vis. (2013) \cite{durrleman_IJCV_2013} & Time warping \\
$\Rightarrow$ Schiratti, NeurIPS (2015) \cite{schirattiNIPS2015}; IPMI (2015) \cite{schiratti_ipmi_2015}; JMLR (2017) \cite{SchirattiJMLR2017} & \\
$\Rightarrow$ Koval, Sci Rep (2021) \cite{Koval2021} & Latent time mixed effects \\
$\Rightarrow$ Poulet, IPMI (2021) \cite{poulet_mixture_2021} & +Subtyping \\
\hline
Donohue, Alz Dem (2014) \cite{Donohue2014} & Latent time fixed effects \\
$\Rightarrow$ Li, Stat Meth Med Res (2017) \cite{li_ltjmm_2019} & Latent-time mixed effects \\ 
\hline
Oxtoby, MICCAI (2014) \cite{oxtoby_bambi_2014} & Differential equation \\ 
$\Rightarrow$ Oxtoby, Brain (2018) \cite{Oxtoby2018} & +Nonparametric \\ 
\hline
Guerrero, NIMG (2016) \cite{guerrero_instantiated_2016} & Instantiated mixed effects \\ 
\hline
Leoutsakos, JPAD (2016) \cite{leoutsakos_alzheimers_2016} & Item Response Theory \\ 
\hline
Lorenzi, NIMG (2017) \cite{lorenzi_gppm_2019} & Nonparametric latent time \\ 
$\Rightarrow$ Garbarino, IPMI (2019) \cite{garbarino_ipmi_2019} & +Differential equation \\ 
$\Rightarrow$ Garbarino, NIMG (2021) \cite{garbarino_GPPM-DS_2021} & Formalism \\ 
\hline
Marinescu, NIMG (2019) \cite{marinescu_dive_2019} & Spatial clustering (c.f. Schiratti/Bilgel) \\ 
\hline
Petrella, Comp. Math. Meth. Med. (2019) \cite{petrella_computational_2019} & Differential equation \\
\hline
Abi Nader, Brain Commun. (2021) \cite{abi_nader_simulating_2021} & Differential equation \\
\hline
Chen, AAAI (2022) \cite{sublign} & Subtyping \\\hline
\end{tabular}
\end{center}
\label{default}
\end{table}%

\begin{nicebox}[Example open source \DPM{} code]
    \label{box:code}
    \begin{itemize}
        \item \DPM{} tutorials: \\\url{https://disease-progression-modelling.github.io}
        \item EuroPOND Software Toolbox: \\\url{https://europond.github.io/europond-software}
        \item KDE EBM: \\\url{https://ucl-pond.github.io/kde_ebm}
        \item pyEBM: \\\url{https://github.com/88vikram/pyebm}
        \item leaspy: \\\url{https://gitlab.com/icm-institute/aramislab/leaspy}
        \item LTJMM: \\\url{https://bitbucket.org/mdonohue/ltjmm} \\\url{https://github.com/mcdonohue/rstanarm}
        \item DPS: \\\href{https://www.nitrc.org/projects/progscore}{source code}; \href{https://hub.docker.com/r/bilgelm/bayesian-ps-adni/}{docker image}
        \item pySuStaIn: \\\url{https://ucl-pond.github.io/pySuStaIn}
        \item TADPOLE-SHARE (from \href{https://tadpole.grand-challenge.org}{TADPOLE Challenge} \cite{marinescu_TADPOLE_2018,marinescu_TADPOLE_2021-short}): \\\url{https://github.com/tadpole-share/tadpole-algorithms}
    \end{itemize}
\end{nicebox}

% References section
\bibliographystyle{spbasicsort}
\bibliography{references}

\end{document}